\documentclass[10pt,conference]{IEEEtran}
\usepackage{cite}
\usepackage{amsmath,amssymb,amsfonts}
\usepackage{algorithmic}
\usepackage{graphicx}
\usepackage{textcomp}
\usepackage{xcolor}
\usepackage[hyphens]{url}
\usepackage{booktabs}
\usepackage{pifont}
\usepackage{siunitx}
\usepackage{amsmath}
\usepackage{algorithm2e}
\usepackage{multirow}
\usepackage{fancyhdr}
\usepackage{float}
\usepackage{setspace}
\usepackage{listings}
\usepackage{balance}
\usepackage{wrapfig}
\usepackage{marginnote}
\usepackage{caption}
\usepackage{floatflt}
\definecolor{gfored}{rgb}{0.580, 0.050, 0.211}
\definecolor{ao}{rgb}{0.007, 0.520, 0.867}
\definecolor{yt}{rgb}{0.875, 0.568, 1.000}
\definecolor{moegi}{rgb}{0.357, 0.537, 0.188}
\definecolor{jl}{rgb}{1.0, 0.2, 0.8}
\definecolor{brown(web)}{rgb}{0.65, 0.16, 0.16}
\definecolor{bisque}{rgb}{1.0, 0.89, 0.77}

\newif\ifsqueezefigs
\squeezefigstrue

\ifsqueezefigs
\makeatletter
\g@addto@macro{\normalsize}{%
  \setlength{\abovedisplayskip}{2pt plus 1pt minus 1pt}
  \setlength{\belowdisplayskip}{2pt plus 1pt minus 1pt}
  \setlength{\abovedisplayshortskip}{0pt}
  \setlength{\belowdisplayshortskip}{0pt}
  \setlength{\intextsep}{2pt plus 1pt minus 1pt}
  \setlength{\textfloatsep}{3pt plus 1pt minus 1pt}
  \setlength{\dbltextfloatsep}{3pt plus 1pt minus 1pt}
  \setlength{\skip\footins}{4pt plus 1pt minus 1pt}}
\makeatother
\fi

\newif\ifdraft
\draftfalse

\ifdraft
    \usepackage[colorinlistoftodos,prependcaption,textsize=small]{todonotes}
    \paperwidth=\dimexpr \paperwidth + 4cm\relax
    \oddsidemargin=\dimexpr\oddsidemargin + 2cm\relax
    \evensidemargin=\dimexpr\evensidemargin + 2cm\relax
    \marginparwidth=\dimexpr \marginparwidth + 2cm\relax

    \newcommand{\ominline}[1]{\textcolor{red}{\textbf{[@om: }#1\textbf{]}}}
    \newcommand{\ombox}[1]{\todo[size=\scriptsize, linecolor=red, bordercolor=red, backgroundcolor=white]{\textcolor{red}{\textbf{@om:} #1}}}

    \newcommand{\agycomment}[1]{\todo[size=\scriptsize, linecolor=orange, bordercolor=orange, backgroundcolor=white]{\textcolor{gfored}{\textbf{@gy:} #1}}}
    \newcommand{\agyinline}[1]{\textcolor{gfored}{\textbf{[@agy: }#1\textbf{]}}}

    \newcommand{\atbcomment}[1]{\todo[size=\scriptsize, linecolor=orange, bordercolor=orange, backgroundcolor=white]{\textcolor{ao}{\textbf{@atb:} #1}}}

    \newcommand{\hluoinline}[1]{\textcolor{moegi}{\textbf{[@hluo: }#1\textbf{]}}}
    \newcommand{\hluobox}[1]{\todo[size=\scriptsize, linecolor=orange, bordercolor=orange, backgroundcolor=white]{\textcolor{moegi}{\textbf{@hluo:} #1}}}

    \newcommand{\yctcomment}[1]{\todo[size=\scriptsize, linecolor=orange, bordercolor=orange, backgroundcolor=white]{\textcolor{yt}{\textbf{@yct:} #1}}}

    \newcommand{\joel}[1]{\textcolor{jl}{#1}}
    \newcommand{\joelcomment}[1]{\todo[size=\scriptsize,linecolor=orange,bordercolor=orange,backgroundcolor=white]{\textcolor{jl}{\textbf{@joel:} #1}}}

\else

    \newcommand{\ominline}[1]{}
    \newcommand{\ombox}[1]{}

    \newcommand{\agycomment}[1]{}
    \newcommand{\agyinline}[1]{}

    \newcommand{\atbcomment}[1]{}

    \newcommand{\hluoinline}[1]{}
    \newcommand{\hluobox}[1]{}

    \newcommand{\yctcomment}[1]{}

    \newcommand{\joel}[1]{{#1}}
    \newcommand{\joelcomment}[1]{}

\fi

\newif\ifrebuttal
\rebuttaltrue

\ifrebuttal
\usepackage[colorinlistoftodos,prependcaption,textsize=small]{todonotes}

\definecolor{darkred}{rgb}{0.9, 0.0, 0.0}

\definecolor{darkblue}{rgb}{0.0, 0.0, 0.85}

\fi

\newcommand*\DRAMCMD[1]{\texttt{#1}}
\newcommand*\DRAMTIMING[1]{t\textsubscript{#1}}

\newcounter{obs}
\newcounter{tkw}
\usepackage{glossaries}

\newacronym{vdd}{$V_{DD}$}{supply voltage}
\newacronym{vpp}{$V_{PP}$}{wordline voltage}
\newacronym{vwl}{$V_{PP}$}{wordline voltage}
\newacronym{vgs}{$V_{GS}$}{gate-to-source voltage}
\newacronym{vth}{$V_{TH}$}{the voltage threshold that the bitline voltage should exceed for the activation to be reliably completed}
\newacronym{gnd}{$GND$}{ground}

\newacronym{ber}{$BER$}{the fraction of DRAM cells that experience bitflips in a DRAM row}
\newacronym{acmin}{$AC_{min}$}{the minimum number of total aggressor row activations to cause at least one bitflip}
\newacronym{ac}{$AC$}{activation count}
\newacronym{rblast}{$r_{Blast}$}{blast radius}
\newacronym{iqr}{$IQR$}{interquartile range}

\newacronym{trcd}{\DRAMTIMING{RCD}}{
{the minimum time between opening a row with an \DRAMCMD{ACT} command and accessing the row buffer}
}
\newacronym{trp}{\DRAMTIMING{RP}}{
{the minimum time between sending a \DRAMCMD{PRE} command and opening a row with an \DRAMCMD{ACT} command}
}
\newacronym{tras}{\DRAMTIMING{RAS}}{
{the minimum time between opening a row with an \DRAMCMD{ACT} command and closing the row with a \DRAMCMD{PRE} command}
}
\newacronym{trefi}{\DRAMTIMING{REFI}}{the \joel{default} time interval \joel{between consecutive \DRAMCMD{REF} commands}}
\newacronym{trefw}{\DRAMTIMING{REFW}}{the maximum time window between two consecutive refresh operations targeting {the same} row}

\def\BibTeX{{\rm B\kern-.05em{\sc i\kern-.025em b}\kern-.08em
    T\kern-.1667em\lower.7ex\hbox{E}\kern-.125emX}}

\author{
{Onur Mutlu}%
\vspace{-3pt}
\\
\emph{ETH Z{\"u}rich}%
\vspace{-20pt}
}

\title{\LARGE{\emph{Retrospective:} RAIDR: Retention-Aware Intelligent DRAM Refresh\vspace{-10pt}}}

\begin{document}
\maketitle
\thispagestyle{plain}
\pagestyle{plain}
\setstretch{0.8}
\begin{abstract}

Dynamic Random Access Memory (DRAM) is the prevalent memory technology used to build main memory systems of almost all computers. A fundamental shortcoming of DRAM is the need to refresh memory cells to keep stored data intact. DRAM refresh consumes energy and degrades performance. It is also a technology scaling challenge as its negative effects become worse as DRAM cell size reduces and DRAM chip capacity increases. 

Our ISCA 2012 paper, RAIDR~\cite{raidr}, examines the DRAM refresh problem from a modern computing systems perspective, demonstrating its projected impact on systems with higher-capacity DRAM chips expected to be manufactured in the future. It proposes and evaluates a simple and low-cost solution that greatly reduces the performance \& energy overheads of refresh by exploiting variation in data retention times across DRAM rows. The key idea is to group the DRAM rows into bins in terms of their minimum data retention times, store the bins in low-cost Bloom filters, and refresh rows in different bins at different rates. Evaluations in our paper (and later works) show that the idea greatly improves performance \& energy efficiency and its benefits increase with DRAM chip capacity. The paper embodies an approach we have termed {\em system-DRAM co-design}.

This short retrospective provides a brief analysis of our RAIDR paper and its impact. We briefly describe the mindset and circumstances that led to our focus on the DRAM refresh problem and RAIDR's development, discuss later works that provided improved analyses and solutions, and make some educated guesses on what the future may bring on the DRAM refresh problem (and more generally in DRAM technology scaling).  

\end{abstract}

\section{Background, Approach \& Mindset}
\vspace{-4pt}

At the time we began our focus on solving the DRAM refresh (i.e., data retention) challenge in late 2010, my research group, SAFARI, had already been working on memory controllers and memory technology scaling issues, motivated by many challenges memory systems, in particular the DRAM technology~\cite{dennard1968field}, have been facing (as described in, e.g.,~\cite{mandelman.ibmjrd02,mutlu.imw13,mutlu.superfri15}). Our intense work on memory systems started during my tenure at Microsoft Research from 2006 and continued at CMU from 2009. For example, we had developed better memory schedulers for multi-core processors (e.g.,~\cite{mutlu-micro2007, mutlu-isca2008, kim2010thread, kim2010atlas, muralidhara2011reducing}), developed platforms to perform voltage and frequency scaling to save DRAM energy (e.g.,~\cite{david-icac11}) and architected emerging memory technologies to replace or augment DRAM (e.g.,~\cite{lee-isca2009, yoon2012row, meza2012enabling}).  We were quite excited about the prospect of much more capable memory controllers in enabling better memory systems. As such, we were pursuing new memory-controller and system-level techniques to 1) overcome the challenging device- and circuit-level scaling issues of memory technologies and 2) better exploit underlying characteristics of memory technology; an approach we termed {\em system-DRAM co-design}~\cite{mutlu.imw13,mutlu.superfri15}. 

RAIDR is a product of this approach. Our focus on data retention issues and other low-level issues in DRAM especially increased via discussions with the Samsung DRAM Design Team, who visited us in April 2011 and encouraged the development of our system-level solutions to DRAM issues, enabling strong support both technically and funding-wise. In fact, much of our ensuing research in DRAM was supported by generous gift funding by and technical discussions with Samsung based on a proposal entitled {\em "New ideas to enhance DRAM scaling: Scaling-aware controller design and co-design of DRAM and controllers"} (Intel provided similar gift funding and technical discussions). 

\section{Contributions and Impact of RAIDR}
\vspace{-3pt}

RAIDR is the first work to propose a low-cost memory controller technique that reduces refresh operations by exploiting variation in data retention times across DRAM rows. Its appeal comes from its simplicity and low cost, enabled by the careful use of Bloom filters~\cite{Bloom:1970:STH:362686.362692}. Exploiting the DRAM data retention time distribution~\cite{dram-retention-2}, RAIDR can eliminate a very large fraction (e.g., $\sim$75\% or more) of refresh operations with very small hardware cost at the memory controller.

Apart from the new technique it introduced, we believe the RAIDR paper made two other major contributions that have enabled a large number of future works and new ideas. First, it provided an empirical scaling analysis that clearly demonstrated the importance of the DRAM refresh problem in modern systems: if nothing is done about it, DRAM refresh would waste almost half of the throughput and half of the energy of a high-capacity 64-Gb DRAM chip!  This analytical prediction encouraged more works in the topic area. Second, it demonstrated a methodical way of exploiting cell-level heterogeneous data retention times at the system (e.g., memory controller) level: if data retention times of DRAM rows are accurately known, the system can use them to optimize DRAM refresh and get rid of most refresh operations. This demonstration enabled other works to develop 1) methods for accurately determining DRAM data retention times and 2) other system-level approaches to optimize DRAM behavior using data retention time information. 

\section{Building on RAIDR and Making It Work}
\vspace{-3pt}

We believe RAIDR enabled a refreshing approach to DRAM refresh. Its largest contribution could be the works it has inspired that rigorously examined the questions of 1) how to perform accurate DRAM data retention time profiling, 2) how to overcome potential hurdles that stand in the way of obtaining accurate minimum data retention times,  3) how to reliably get rid of unnecessary refresh operations.  

We wanted to make RAIDR work in a real system setting. To this end, collaboratively with Intel, we developed an FPGA-based flexible DRAM testing infrastructure~\cite{liu.isca13} that enabled us to rigorously test data retention times of cells in real DDR3 DRAM chips. Using this infrastructure, later open sourced as SoftMC~\cite{hassan2017softmc, softmc.github} and DRAM Bender~\cite{olgun2023drambender, safari-drambender}, we experimentally examined practical issues that affect the accuracy (and performance) of DRAM data retention time profiling. We analyzed two major issues that make such profiling very challenging: 1) data pattern dependence (DPD) of retention times~\cite{data-line-interference,liu.isca13}, and 2) the variable retention time (VRT) phenomenon~\cite{yaney1987meta, restle1992dram,liu.isca13}. Our follow-up work, which appeared at ISCA 2013~\cite{liu.isca13}, provides a detailed experimental analysis of these challenges in cutting-edge DRAM chips, demonstrating that ideas like RAIDR that depend on accurate identification of retention times are not easy to exploit in practice. Later works (e.g.,~\cite{khan.sigmetrics14, avatar-dsn15, khan.dsn16, khan.cal16, khan.micro17, patel2017reaper, patel2020beer, patel2021harp}) developed new methods for making RAIDR-like techniques more practical by tackling especially the DPD and VRT problems and enhancing retention time profiling methods to work in the presence of DPD and VRT, usually by exploiting ECC techniques that have since become mainstream in DRAM chips (see~\cite{patel2019understanding, patel2020beer, patel2021harp}) to tolerate VRT~\cite{kang2014co}. 

The development of our flexible FPGA-based DRAM testing infrastructure also enabled experimental DRAM research in directions that are completely different from retention time profiling and refresh. These include studies that provided valuable experimental data on various DRAM characteristics, including RowHammer~\cite{kim_flipping_2014, kim2020revisiting, orosa2021deeper, farmani2021rhat, frigo2020trr, hassan2021utrr, yaglikci2022hira, yaglikci2022understanding, olgun2023drambender, olgun2023hbm, luo2023rowpress}, latency~\cite{lee.hpca15, chang.sigmetrics16, lee2017design, kim2018solar}, voltage-latency-reliability relationship~\cite{chang.sigmetrics17}, power consumption and modeling~\cite{ghose2018vampire}. Using this infrastructure, later research also demonstrated the ability of real off-the-shelf DRAM chips to perform data copy/initialization and bulk bitwise operations~\cite{Seshadri:2015:ANDOR, seshadri_ambit_2017, pidram, gao2019computedram, gao2022fracdram}, implement physical unclonable functions~\cite{kim2018dram}, and generate true random numbers~\cite{kim.hpca19,olgun2021quactrng}. We believe the investment we  made to try to make RAIDR work using a real FPGA-based infrastructure helped us and the broader research community uncover many interesting  characteristics of DRAM chips and propose new ideas to make DRAM-based systems more secure, reliable, efficient, and high performance. 

Other later works provided refined models of DRAM refresh's impact on system performance (e.g.,~\cite{chang2014improving, mukundan2013understanding}) and developed new methods to reduce DRAM refresh's negative impact on performance \& energy (e.g.,~\cite{lin2012secret, nair2013arch, chang2014improving, mukundan2013understanding, nair2013case, zhang2014cream, hassan2019crow, hassan2022case, das2018vrldram, yaglikci2022hira}). Our HPCA 2014 paper~\cite{chang2014improving} developed a more refined projection of the effect of DRAM refresh as technology scales. AVATAR in DSN 2015~\cite{avatar-dsn15} and REAPER in ISCA 2017~\cite{patel2017reaper} enabled more practical ways of exploiting heterogeneous retention times in the presence of VRT. Our recent work~\cite{hassan2022case} shows that with a more flexible DRAM interface that gives some autonomy to DRAM chips, RAIDR can be more efficiently implemented inside the DRAM chip.

\section{Summary and Future Outlook}
\vspace{-3pt}

RAIDR is a nice example of how enthusiastic support from industry can foster new ideas that can open up many new analyses and other ideas. We were inspired by our deep technical discussions with especially Samsung and Intel, along with prior works that described DRAM technology scaling challenges (e.g.,~\cite{mandelman.ibmjrd02}) and that developed promising solutions (e.g.,~\cite{venkatesan2006rapid, song2011flikker}). Engineers from Samsung and Intel later wrote an insightful paper~\cite{kang2014co} on DRAM scaling challenges, which described refresh as a key problem and advocated a controller-DRAM co-design approach as we had been advocating~\cite{mutlu.imw13, raidr}. RAIDR was also a nice example of how teaching \& research smoothly feed each other: much of the research was done as part of a group project in the Parallel Computer Architecture class I taught at CMU in Fall 2011.

Looking forward, DRAM technology scaling is getting worse and data retention will continue to be an important issue~\cite{kang2014co, sk-hynix-isscc2023}. The negative effects of DRAM refresh will be (and are being) exacerbated by other technology scaling issues like RowHammer~\cite{kim_flipping_2014} that require even more refreshes as a solution~\cite{mutlu2023fundamentally, yaglikci2022hira, luo2023rowpress}. We believe there are a lot more new ideas and techniques to develop to minimize the impact of refresh on computing systems.

\setstretch{0.70}

\balance
\bibliographystyle{IEEEtran}
{\tiny
\bibliography{combined}}

\end{document}